\documentclass[final,5p,times,english,twocolumn]{elsarticle}

\usepackage{amssymb}
\usepackage{tensor}

\usepackage{amsmath}
\usepackage{xcolor}
\usepackage{subcaption}
\usepackage{dsfont}
\usepackage{mhchem}
\usepackage{background}

\journal{arXiv}

\usepackage{hyperref}
\hypersetup{
     colorlinks=true,
      linkcolor=blue,
}

\DeclareMathOperator{\arsinh}{arsinh}
\DeclareMathOperator{\erf}{erf}

\backgroundsetup{contents=}
\begin{document}

\begin{frontmatter}



\title{Comments on the comments by Lackner et al. on the series of
  papers about ``A novel direct drive ultra-fast heating concept for ICF''}


\author{Hartmut Ruhl and Georg Korn} 

\address{Marvel Fusion, Theresienh\"ohe 12, 80339 Munich, Germany}

\begin{abstract}
In this paper, we provide a response to the comments made by Lackner
et al. regarding our series of recent papers on "A novel direct drive
ultra-fast heating concept for ICF". Specifically, we comment on the
necessity of fuel pre-compression in the ICF context.
\end{abstract}


\end{frontmatter}


\tableofcontents

\section{Introduction}
In a series of papers
\cite{ruhlkornarXiv,ruhlkornarXiv1,ruhlkornarXiv2,ruhlkornarXiv3}, we
have been investigating the feasibility of a new variation of an
inertial confinement fusion (ICF) reactor concept. Our approach
involves the use of short laser pulses, nanostructures, and
alternative mixed fuel types, aiming to expand the range of potential
ICF variants with much reduced technological complexity.

Ultra-short laser pulses, when interacting with structured foams or
nanostructures, have the remarkable capability of delivering both
ultra-high power and a substantial amount of energy to a target
by fast ions, electrons, and radiation simultaneously. In our specific
context, we want to highlight their potential to achieve fuel
temperatures high enough to initiate ignition in a new category of
fuels we refer to as 'mixed fuels' in our research papers.

It's worth noting that the primary motivation for pre-compressing the
traditional $\ce{DT}$ fuel in the ICF context is to enhance the
stopping power of $\alpha$-particles within the $\ce{DT}$ fuel mass to
an extent that the fuel mass is manageable while achieving sufficient
fusion gain at the same time. This implies a reduction of the required
external energy for fuel heating as well. Mixed fuels, on the
other hand, tend to exhibit significantly enhanced stopping power
compared to $\ce{DT}$ at their natural densities, effectively
simulating compression in $\ce{DT}$. The technological challenge to
our believe lies in successfully igniting mixed fuels on very short
time scales and in confining their energy sufficiently long while fuel
pre-compression is avoided. Avoiding fuel pre-compression, however,
by no means excludes fuel compression. Such technologies may have come
into existence recently.

Lackner et al. make several comments that are based on presumptions
about the focus of our papers and the IFE-strategy of the Marvel
Fusion. However, these presumptions do not align with the actual focus
and intent of our work. Our focus is on structured foams and our
intent is purely academic. Our papers do not represent the IFE-strategy
of Marvel Fusion and to our knowledge such a strategy has not been
published. However, it is fair to say that Marvel Fusion focuses on
efficient repetition-rated broadband laser technology, which is downward
compatible with the HYPER laser project. In other words, the
prototypical laser technology Marvel Fusion has is capable of emulating
HYPER. As a consequence, almost all classical high gain fusion target
concepts discussed in the ICF community based on deeply engrained
believes are accessible, while at the same time the broadband
technology favored by Marvel Fusion offers upside options in the realm
of ultra-short laser pulses at no substantial extra cost. To which
extent high power density driver concepts widen the scope of ICF
fusion concepts is the subject of our ongoing open source arXiv
project. At present, however, we focus on the fuel heating power of
structured foams or in other words ordered nanostructures.

In this paper, we provide a comprehensive response to the comments
made by Lackner et al. in their publications, namely \cite{karlchenvomIPP3, karlchenvomIPP5,
  karlchenvomIPP4, karlchenvomIPP1}, which are relevant to our recent
series of papers \cite{ruhlkornarXiv,ruhlkornarXiv1,ruhlkornarXiv2,ruhlkornarXiv3}.
We would like to express our gratitude for the hard work and critical
comments provided by Lackner et al., which we address and discuss in
the present paper to the extent they are relevant.

\section{Our comments on \cite{karlchenvomIPP3}}
In our paper \cite{ruhlkornarXiv2}, a typographical error occurred in
the fractional fuel densities. The values for the fuel densities
considered in the paper are $n_p = n_D = n_T = 5 \, n_B/3$. Next, we
would like to address an inaccurate claim made by Lackner et
al. \cite{karlchenvomIPP3} in their comments
\begin{itemize}
\item {The paper by Marvel Fusion utilizes the energy of
    neutrons to estimate the fusion energy that can be deposited in
    the fuel.}
\end{itemize}
We would like to clarify that in our study \cite{ruhlkornarXiv2}, we
do not take into account neutrons for in-situ fusion energy deposition
in the fuel. The inaccuracy of their claim could have been deduced
from Figure 5 in our paper, which clearly demonstrates that no
neutrons are involved.

\section{Our comments on
  \cite{karlchenvomIPP5,karlchenvomIPP4,karlchenvomIPP1}}
Lackner et al. raise three essential concerns:
\begin{itemize}
\item[1.] {Compression is always needed and absolutely essential}.
\item[2.] {The time scales required for nuclear fusion and
    those of femtosecond lasers differ too much}.
\item[3.] {The short duration of the laser - nanostructure
    coupling limits the acceleration phase and hence the total
    energy coupled into the fuel and ablated mass}.
\end{itemize}
Response to bullet point 1: The main reason for the
need of fuel pre-compression in $\ce{DT}$ is reducing the range of the
$\alpha$-particles such that they stop at a fuel mass small enough to
be mangeable. Mixed fuel types are different to this extent. Their
$\alpha$-particle stopping power is much higher potentially implying
small enough fuel mass for manageable energy production at their
natural densities. While $\ce{DT}$ requires fuel pre-compression it
tends to ignite at low temperatures. Mixed fuel types require much
higher fuel temperatures. We propose to reach them with the help of
structured foams. As we state in our papers fuel compression tends to
reduce the amount of the required external energy in all fuels for
arbitrary gain requirements. However, we do not consider fuel
pre-compression in our papers at present.

Response to bullet point 2: Lackner et al.'s second claim is
erroneous. Ultra-intense ultra-short laser pulses can contrary to
the claim have long-lasting effects on fluels. One notable example is the
potential capability of ultra-fast volumetric fuel heating by short-pulse lasers
via rapidly accelerated particles and radiation beyond the ignition
point. To which extent ultra-high radiation densities can compress
fuel we choose not to address at present.

Response to bullet point 3: Lackner et al.'s assertion, as
articulated in reference \cite{karlchenvomIPP5}, that short-pulse
laser-accelerated ions cannot induce compression due to their limited
rocket effect, is not correct. The concept of impact fusion challenges this
assertion. By employing fast fluids characterized by high mass
density, it becomes feasible to compress and ignite fluids with lower
mass density. In this approach, assuming the heavier masses can be rapidly
accelerated within an ultra-short timeframe using the intense
polarization fields inherent in nanorods, there is the potential to attain
substantial fuel compression through the impact of the heavier
fluid component on the lighter one. However, mixed fuels may not
require fuel pre-compression in the first place.

\section{Summary}
Lackner et al. put forth three key claims: i) achieving high gain
fusion is not possible without compression, ii) femtosecond lasers are
unsuitable for problems that operate on hydrodynamic time scales, and
iii) ions accelerated by short-pulse lasers are incapable of exerting
a sufficiently strong and long-lasting rocket effect to influence fuel flows.

In response to i), we assert that not only accelerating energy
production in the fuel by higher fuel density but also the stopping
power and with it the degree of in-situ fusion energy feedback as well
a slowing down other loss processes are crucial factors affecting fusion
gain. We believe that mixed fuels hardly require fuel pre-compression
and can rely on other parameters that play a role in achieving
efficient fusion.

Regarding ii), we hope to have made the case that making a
general statement about the inadequacy of short pulse lasers for fusion
due to the mismatch between hydro time scales and femtosecond laser
pulse lengths cannot be made. The effectiveness of femtoseconds laser
pulse technology for a particular fuel depends on various
factors and cannot be generalized without a thorough evaluation.

In response to iii), we mention the concept of impact fusion, which is
actively discussed in the ICF community. Impact fusion does not
necessarily rely on a rocket effect. It is premature to dismiss the
effectiveness of short-pulse lasers interacting with nanorods for
the sustained impact necessary for achieving high gain.

Lastly, it is worth noting that simulations with the help of an ICF
community code indicate that setting up an uniform temperature profile
in a mixed fuel as might be possible by nanorods can lead to $Q_T > 1$
at reasonable external energy levels \cite{ruhlkornarXiv3}. There are
numerous ways to enhance gain. At present they are not in our focus
since the nanostructures need to prove their proposed capability in the
experiments first. Furthermore, low gain ICF concepts for neutron
producing via mixed fuels may have applications for and
beyond energy production.

\bibliographystyle{elsarticle-num} 
\bibliography{literatur_eqn_motion}

\end{document}